\documentclass[showpacs,prl]{revtex4} 
\usepackage{graphicx}
\usepackage{bm}
\begin{document} 

\title{Differential Effect of Delays and Network Topology on Temporal Pattern Formation}

\author{Rhonda Dzakpasu and Micha{\l}\ \.Zochowski} 
\affiliation{Department of Physics and Biophysics Research Division\\
University of Michigan\\
Ann Arbor, MI 48109}

\begin{abstract}
 We investigate the effects of transmission delays on the formation of temporally ordered states in networks of non-identical R\"ossler oscillators, having SWN topology.  We show that incorporation of two different types of delay, length independent and length dependent, leads to dramatically different temporal properties of the network. In the first case formation of global random connections leads to increased temporal ordering, while in the second case it annihilates locally ordered clusters and forms a disordered state. 

\end{abstract} 
\date{\today}
\maketitle 
Studies focused on elucidating properties of networks having different topologies have generated a lot of interest. These studies span diverse systems ranging from computer \cite{balthrop2004}, social \cite{porter2005,meyers2005,newman02} and citation \cite{bilke2001,henrickson2004} networks to biological ones such as neural networks \cite{percha_mrz2005,netoff2004,lago00}. Emergence of the concept of small world networks (SWN) \cite{strogatz} allows for the rigorous study of the properties of intermediate structured networks, i.e., having structure between local connectivity and random graph, as a function of their topology. Those studies have mostly concentrated around the static properties such as characteristic path length and clustering coefficient \cite{newman00,almass02}, scaling and percolation properties of such systems \cite{newman99,newman02} as well as possible mechanisms underlying their formation. 

Recently, however, more attention has focused on the formation of spatio-temporal patterning in networks having different topologies. It has been shown that the linear stability of the synchronous state is linked to the algebraic condition of the Laplacian matrix defining network topology \cite{barahona02,hong02} i.e., the synchronized state is stable if the blocks of the matrix that are associated with the graph eigenmodes transverse to synchronization manifold have negative Lyapunov exponents \cite{barahona02}. In general, it has been reported that this synchronized state is achieved in SWN more efficiently in terms of required network connectivity than standard deterministic graphs, purely random graphs and ideal constructive schemes \cite{nishikawa03}. In biological networks, it has been shown that small-world networks of interconnected Hodgkin-Huxley neurons combine two features: rapid and large oscillatory response to the stimulus \cite{lago00}. It has been also hypothesized based on SWN paradigm that the increased rewiring of the networks possibly observed during axonal sprouting could cause formation of large scale pathological synchrony - epileptic seizures \cite{percha_mrz2005,netoff2004}. 

Those studies however assumed that there is no conduction delay between the network nodes or that the delay is intrinsic to the connection itself and does not depend on the geometrical properties of the space in which the network is immersed, i.e., Euclidean distance between the two nodes.  

In this letter we investigate temporal patterning in a 2D network of non-identical R\"ossler oscillators, having SWN topology and two types of delay: length independent delay and connection length dependent delay. The first case pertains to situations where the conduction delays in a wire (connection) are small but every node relay imposes the delay in the transmission, whereas the second case describes the situation where the network sprawls or the transmission speed is not constant. The network topology will then have to incorporate varying network delays that scale with Euclidean distance and/or speed of transmision. Neural networks are a good example of both types of network, as at short distances axon transmission is fast enough to be neglible and (small) delays are generated at chemical synapses themselves, whereas at long distances the axonal delays play an important role in information processing \cite{crook1997}. We show that inclusion of these two types of delays leads to diametrically different results in terms of temporal behavior of the networks having varied topologies (i.e. different degree of rewiring).

It has been established that periodically-driven non-linear oscillators or a system of coupled non-identical oscillators can achieve phase or, with stronger coupling, time-lag synchronization \cite{rosenblum,rosenblum2,pikovsky,parlitz2,zhou02,pazo03}. Phase synchrony is characterized by the fact that the phases (lags) between the oscillators are locked while their amplitude will vary. It has been shown that the increase in delays between pairs of such oscillators can be offset by the increase of the coupling strength and the system synchronizes even for large delays \cite{chen2002}, while in a network they lead to cluster formation \cite{park1999,choi2000} and may play important role during the information processing in neural systems \cite{crook1997}.

We investigate a network comprised of $N=144$ diffusively coupled, nonidentical R\"ossler oscillators that are positioned at the nodes of a 2D lattice. The 12x12 lattice has periodic boundary conditions, i.e., toroidal topology and a lattice constant $q=1$. The system equations are as follows:
\begin{equation}
\begin{array}{l}
\dot{x}_{i}(t)=-(z_{i}(t)+y_{i}(t))\\
\dot{y}_{i}(t)=x_{i}(t)+a_{i}y_{i}(t)+ \frac{\alpha}{K-1}\sum_{j, j\neq i}^K (y_{j}(t-\tau_j)-y_{i}(t))\\
\dot{z}_{i}(t)=b+(x_{i}(t)-c)z_{i}(t) 
\end{array}
\end{equation}

where $b=0.2$, $c=10.0$ are the oscillator parameters; $\alpha=0.4$ defines the coupling strength between the oscillators. The subscripts denote oscillator number and $K$ is the number of actual connections per oscillator. The parameter, $a_i\in [0.1, 0.35]$ is randomly generated and unique for each oscillator. The variable $\tau_j$ denotes the delay between the coupled nodes. For length independent delays, $\tau$ is constant while for length dependent delays it is proportional to the Euclidean distance between the nodes on the toroidal surface. 

Initially all nodes within the radius $r=1$ are connected, and to modify the network topology a fraction $f$ of the connections is randomly rewired. Thus for $f=0$ the network has only local connections, while for $f=1$ the network is a random graph.  

The effective delay between the nodes varies widely for the length dependent and length independent delays as a function of network topology. In the length independent case the delay between any two nodes regardless of their relative position in the network is constant. Thus, the average effective delay between any two nodes in the network depends only on the characteristic length in the system (i.e., how many connections are needed to travel from one to the other). This characteristic length decreases as the fraction of global random connections gets larger (Fig. 1A). On the other hand, for length dependent delays, the delay between the nodes does not depend on the characteristic length in the network, and moreover the average delay per node as well as its variability increases linearly with the rewired fraction (Fig. 1B).
\begin{figure}
\includegraphics[scale=0.6]{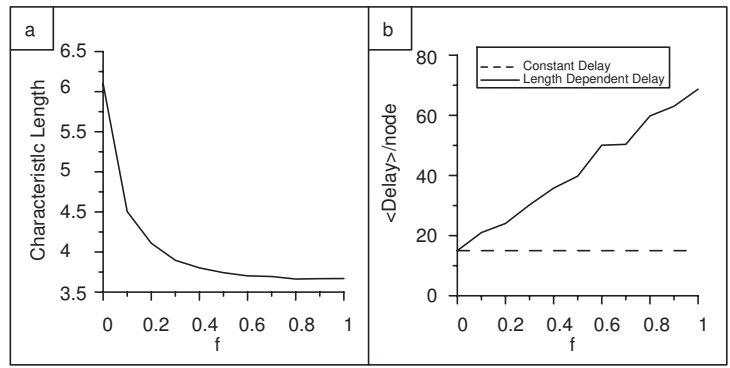}
\caption{Topology and effective network delays. A) Characteristic length in the network as a function of rewiring parameter $f$. An $NxN$ connectivity array $A$ is formed where unity denotes a direct connection between appropriate nodes and zero otherwise.  $A^m$ for $m=1,2,3,..$ is computed. The lowest value of $m$ for which there are no zero elements in the array corresponds to the characteristic length of the network. The curve is an average of 8 random realizations of the network. The delays for the length independent case are directly proportional to the characteristic length while for the length dependent case they depend on the Euclidean distance. B) The average delay per node is calculated for both length independent and length dependent delays. For length independent delays, the average delay is independent of the distance between the nodes and thus is constant. Conversely for length dependent delays, the average delay increases with $f$.}   
\end{figure}

To assess the phase synchrony as well as temporal ordering in the networks incorporating these two types of delays we use an earlier developed measure \cite{mrzrd1-03,dzakpasu2005,percha_mrz2005} that monitors asymmetries in the variability in the lags between the relative interevent intervals (IEI) generated by the oscillators. In the current work, an event is defined as the time when the trajectory of the oscillator crosses a Poincare section $z=0$. Specifically, the inter-event interval, $\Delta t^{ij}_m$, of the $j$-th unit with respect to the $i$-th unit is calculated as a time difference between the event time of the $j$-th oscillator and the last event of the $i$-th unit that directly preceeds it, and conversely the inter-event interval $\Delta t^{ji}_m$ of the $i$-th unit is calculated as a time difference between the event of $i$-th node and the timing of the last event generated by the $j$-th oscillator that directly preceeds it. The IEIs are calculated for every oscillator pair in the network separately as new events are generated. The distributions are updated dynamically throughout the simulation, so that if the inter-event interval created at time $t$ falls within a time window described by bin $I$, the probability assigned to that bin becomes $P_I(t)=P_I(t-1)+ \Delta P$; $\Delta P$ is the free parameter of the measure and determines its dependence on the previous IEIs. Thus, if $t^i_1, t^i_2$ are the event timings generated by $i$-th oscillator, while $t^j_1, t^j_2$, are the timings of the events generated by the $j$-th oscillator, and $t^i_1>t^j_1>t^i_2>t^j_2$, then one distribution will be updated with only one inter-event interval ($\Delta t^{ji}=t^i_2-t^j_1$) whereas the second distribution will be updated twice with ($\Delta t^{ij}=t^j_1-t^i_1$) and ($\Delta t^{ij}=t^j_2-t^i_2$). The distributions defined in this manner are complementary to each other, providing complete information about any possible asymmetric timing interdependencies between the coupled oscillators. After every update, the ISI distributions are renormalized and their Shannon entropy, $S=-\sum_I P_I\log{P_I}$,  is calculated. 

When phase synchrony is achieved, the inter-event intervals calculated from the leading oscillator to the following one will vary much less than the inter-event intervals from the following oscillator to the next event of the leading oscillator, causing in turn the entropy of those distributions to be much lower \cite{mrzrd1-03,dzakpasu2005,percha_mrz2005}.

The lag between two non-identical oscillators established during phase/lag synchronization depends on their relative properties (i.e. intrinsic frequencies) \cite{rosenblum,rosenblum2}. The oscillators with the higher frequency (higher $a_i$) will lead oscillators with a lower frequency. To measure this temporal ordering we define an {\it expectivity} function that compares the temporal interdependencies in the network to the relative properties of the oscillators i.e., the value of parameter $a_i$:
\begin{equation}
E(t)=\frac{1}{N(N-1)}\sum_{i,j\\,i\neq j}^N w_{ij}(t),
\end{equation}   
where
\begin{equation}
w_{ij}(t)=\left\{
\begin{array}{rl}
1 & \mbox{if $(S_{ij}(t)-S_{ji}(t))(a_j-a_i)>0$}\\
-1 & \mbox{if $(S_{ij}(t)-S_{ji}(t))(a_j-a_i)\leq 0$}
\end{array}
\right.
\end{equation} 
The $S_{ij}(t)$ is the entropy calculation described above for the $j$-th oscillator with respect $i$-th one. Their difference ($S_{ij} - S_{ji}$) will provide a direct assessment of the direction of the lag between the two oscillators.
Thus the expectivity function provides a measure of the existence of global temporal ordering in the system: namely, whether the direction of the phase lag is in agreement with that predicted from the relative values of the parameters ($a_i$) of the oscillators. If the value from a given pair is predicted correctly the function is assigned the value $w_{ij}=1$, and conversely if the prediction fails $w_{ij}=-1$. Thus, if $E\rightarrow 1$ this indicates that there is a global order in the temporal sequences of oscillator activities, whereas if no ordering is established $E\simeq 0$. 

We have applied the expectivity to monitor changes in the degree of temporal ordering for networks having both length independent and length dependent delays. Figure 2 depicts the changes in the average expectivity (Fig. 2A,C) and its standard deviation (Fig. 2B, D) over a 300 000 step integration time, as a function of rewiring parameter $f$. Each point on the graph is an average over 8 simulations. For length independent delays the rewiring induces a larger degree of temporal ordering in the network, as the expectivity increases from $E\simeq 0.2$ for $f=0$ to $E\simeq 0.8$ for $f=1$ for small and intermediate values of the delay (Note: the delay $\tau$ is represented as a fraction of the average period of the oscillators in the network; for length dependent delays $\tau$ represents the delay per unit distance). The expectivity never reaches $E=1$ due to the sparse connectivity of the network.
For a larger delay, $\tau\simeq 0.06$, the ordering is initially low (expectivity is near zero) but then also improves rapidly for $f\geq 0.8$. We have shown previously that for a network of neural oscillators with no delays, the rewiring induced a phase transition from a temporally disordered into a globaly ordered state, with the "expectivity" function being the order parameter of the system \cite{percha_mrz2005}.
\begin{figure}
\includegraphics[scale=0.6]{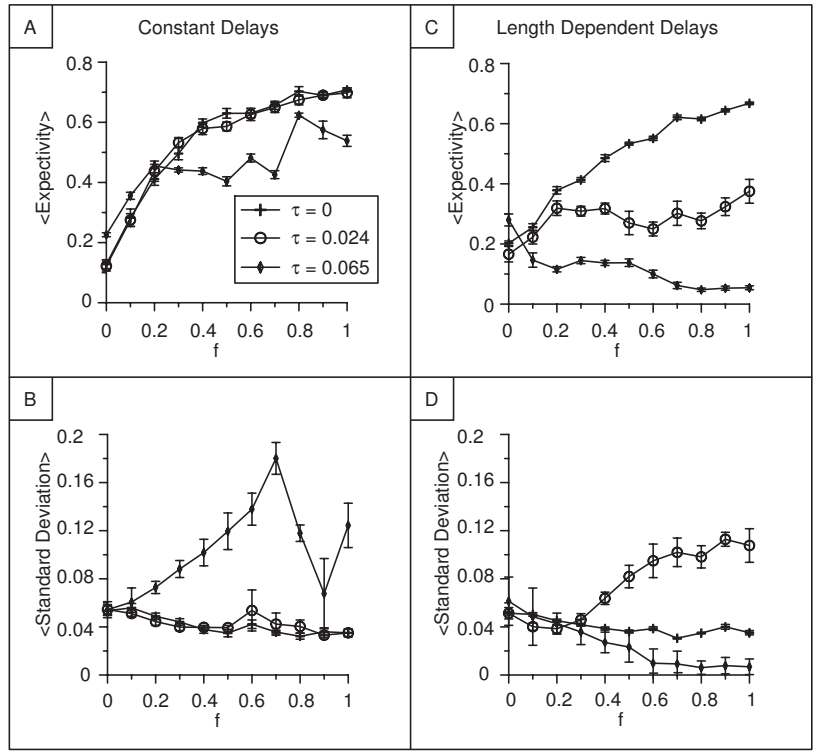}
\caption{Average global expectivities and variability as a function of rewiring fraction $f$.  A,C) Average global expectivities calculated for length independent and length dependent delays, respectively. For length independent delays, the expectivity increases as the network transitions to a random graph, indicating increased ordering in the system. For the length dependent delays, the expectivity decreases for large delays $\tau$ with the rewiring parameter $f$. B,D) Average standard deviations of the global expectivities within each simulation run, calculated for length independent and length dependent delays, respectively. Every point on the graphs is an average of 8 runs. Every run consists of 
$300 000$ iterations.}   
\end{figure}

On the other hand, the behavior of the expectivity for the case of length dependent delays is dramatically different. The increase in rewiring does not increase the expectivity in the network for intermediate values of the delay per unit length. For larger values of the delay, the large fraction of random connections systematically destroys any ordering in the network. Additionally for the intermediate delay the system becomes unstable achieving a relatively large variability in the expectivity during its evolution.

To better understand the changes in network spatio-temporal patterning as a function of its topology, we investigated how the expectivity scales as a function of distance between two nodes in the network. For length independent delays (Fig. 3), for low values of $f$ (Fig. 3A) the expectivity is significantly higher at short distances and decays for longer ones, indicating the formation of local clusters in the network. For large values of $f$, (Fig. 3B-D), a globally synchronized state is gradually formed with a high degree of temporal order. 
\begin{figure}
\includegraphics[scale=0.6]{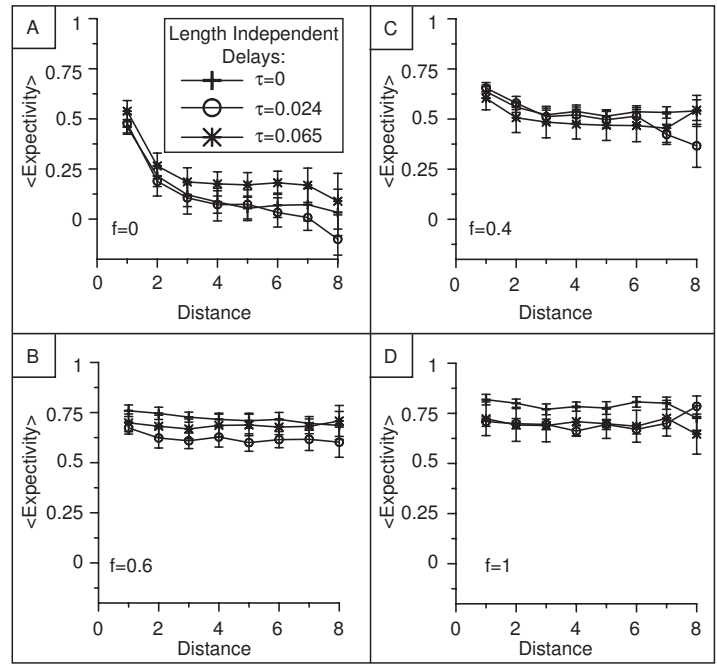}
\caption{Expectivity vs. the distance between the nodes for length independent delays. As the rewiring parameter, $f$, increases to unity, the expectivity increases through out the network.  The local ordering transitions to a globally ordered system. A: $f=0$; B: $f=0.4$; C: $f=0.6$; D: $f=1.0$.}   
\end{figure}
Conversely for the length dependent delays, (Fig. 4), the local clusters that are initially formed (high expectivity over short distances; Fig. 4A) are enhanced or destroyed for random graph topologies depending on the value of the delays in the network (Fig. 4B-D). Thus rewiring in this case may effectively destroy local clusters exhibiting temporal ordering in order to form a globally unordered state.
\begin{figure}
\includegraphics[scale=0.6]{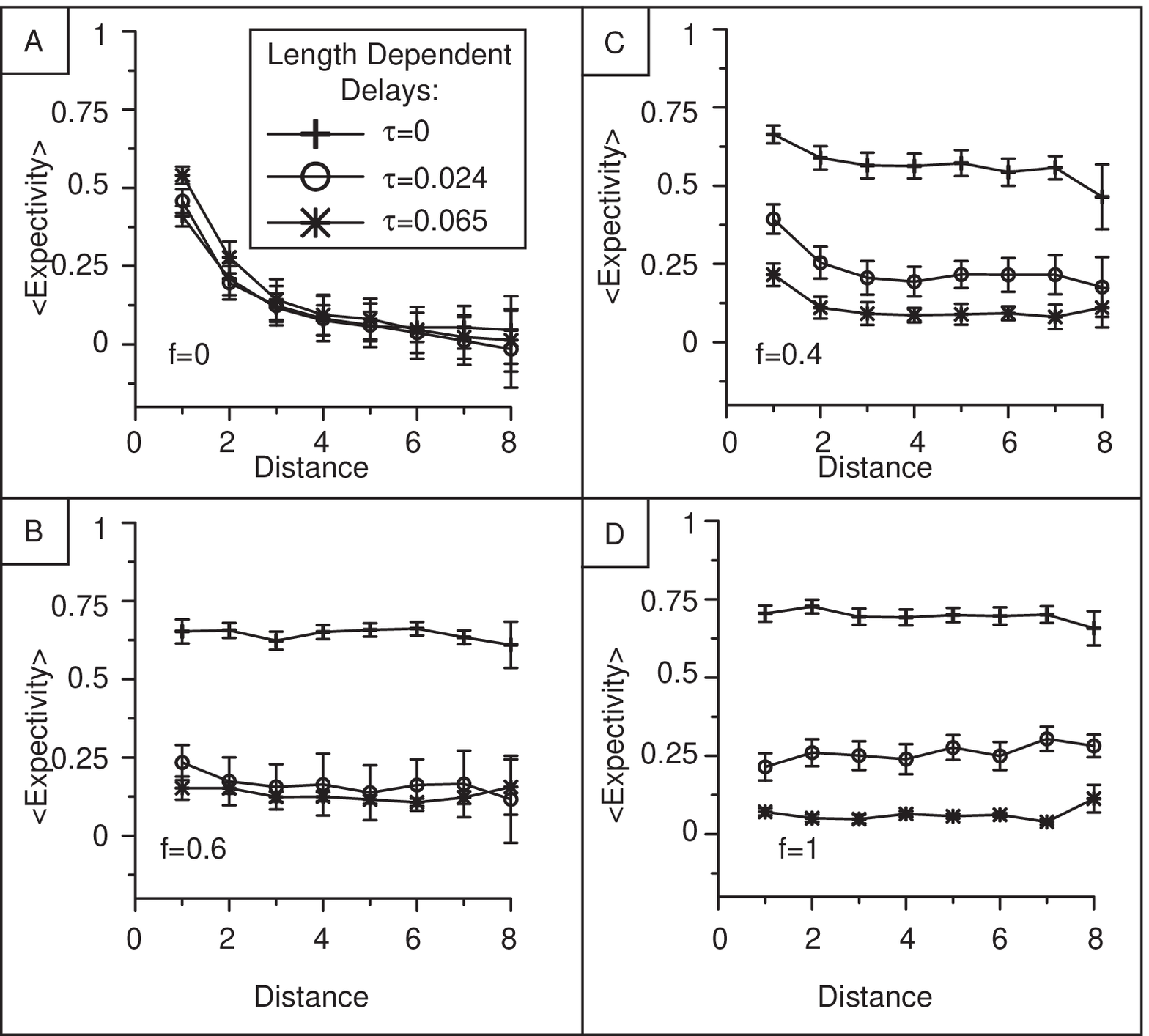}
\caption{Expectivity vs. distance between the nodes for length dependent delays. As the rewiring parameter, $f$, increases to unity, the expectivity decreases to zero for nodes with any separation within the network. The random rewiring acts to destroy local clusters and forms a globally disordered state. A: $f=0$; B: $f=0.4$; C: $f=0.6$; D: $f=1.0$.}   
\end{figure}
To better illustrate this effect we plot the changes of expectivity calculated only for nearest neighbors as a function of the rewiring parameter for length dependent and length independent delays for an intermediate value of $\tau$ (Fig. 5). For length independent delays the average ordering between the neighbors increases with the rewiring parameter to achieve the global value of $E\simeq 0.8$. At the same time the order in the network with length dependent delays is effectively destroyed.    
\begin{figure}
\includegraphics[scale=0.9]{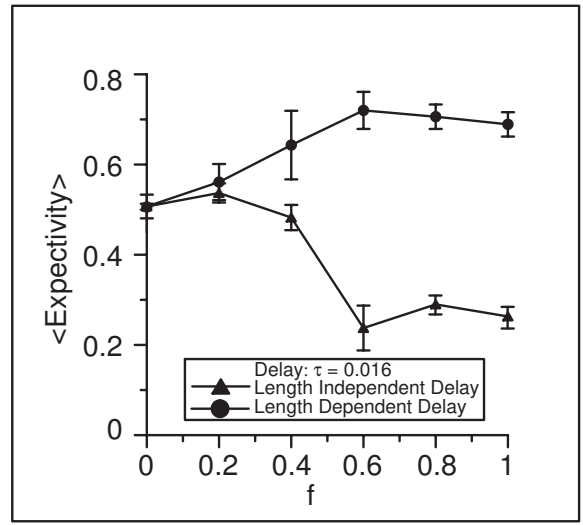}
\caption{Changes of expectivity as a function of rewiring parameter $f$, for neighboring nodes for an intermediate value of delay.
For length independent delays the ordering between nearest neighbors significantly increases, whereas for length dependent delays it rapidly decays for larger fractions of the rewired connections.}   
\end{figure}
In summary, we have investigated a network having different topologies and incorporating two different types of delays - length independent and length dependent. We show that topologically and functionally these two types of delays produce networks with dramatically different properties. In the first case the formation of global random connections limit the effective delay between any two nodes in the network and induces temporal ordering in the system as a whole. Conversely, in the second case, the formation of random global connections effectively increases the average delay per node and creates delay variability. Thus, higher values of the delays cause annihilation of the local temporally ordered cluster and produce a globally disordered state. The formation of a new global connectivity causes the destruction of the temporal ordering in the system. This could be an important mechanism for example in neural systems to facilitate the information processing to limit formation of pathologically synchronous states for example during epilepsy \cite{crook1997}. 

We would like to thank Mark Newman for his comments. This work has been supported by NIBIB grant no. R21EB003583.

\end{document}